\documentclass[conference]{IEEEtran}
\IEEEoverridecommandlockouts

\usepackage{cite}
\usepackage{amsmath,amssymb,amsfonts}
\usepackage{algorithmic}
\usepackage{graphicx}
\usepackage{textcomp}
\usepackage{xcolor}
\usepackage{float}
\usepackage[bottom=1in,top=0.75in, left=0.625in, right=0.625in]{geometry}
\def\BibTeX{{\rm B\kern-.05em{\sc i\kern-.025em b}\kern-.08em
    T\kern-.1667em\lower.7ex\hbox{E}\kern-.125emX}}
\begin{document}

\title{Smart Voltage Monitoring: Centralised and Blockchain-based Decentralised Approach\\
}

\author{\IEEEauthorblockN{Shailesh Mishra}
\IEEEauthorblockA{{IIT Kharagpur, Kharagpur, India} \\    
mshailesh0511@iitkgp.ac.in}
\and
\IEEEauthorblockN{Shivam Kumar}
\IEEEauthorblockA{{IIT Kharagpur, Kharagpur, India} \\
shivamk2000@iitkgp.ac.in}
}

\maketitle

\begin{abstract}
Voltage controls the majority of the processes around us, starting from lighting an incandescent lamp to running huge machines in industries. Therefore, voltage monitoring becomes essential, which demands efficient measurement and storage of voltage data. However, there is hardly any system till date that fulfils both the goals of voltage monitoring and voltage data storage. To achieve this goal, we propose the application of the Internet of Things along with the server-based framework and Distributed Ledger Technology to build systems for smart voltage monitoring. Two models - a centralised model and a decentralised model have been presented and analysed thoroughly in this paper. The centralised model is built on client-server architecture, whereas the decentralised model is based on a peer-to-peer architecture. Blockchain and InterPlanetary File System have been used for the implementation of the decentralised system. Potential improvements to make these systems robust have also been discussed. The methods proposed in this paper for voltage monitoring are novel; ensure efficient data storage and can be used for IoT data storage of any form.
\end{abstract}

\begin{IEEEkeywords}
Internet of Things(IoT), Client-Server, Blockchain, InterPlanetary File System(IPFS), Decentralisation, Voltage
\end{IEEEkeywords}
\section{Introduction}\label{sec:intro}
Electricity has been transforming our lives since the day it was discovered. It has given rise to numerous devices without which our life would be unimaginable. Therefore, their proper operation is necessary for our lives to flow smoothly. The physical quantity that controls this adequate operation is the voltage. Any fluctuation in voltage is detrimental to the operation of all the appliances. Variation in voltage leads to power loss which is harmful both economically and environmentally. Hence, control of voltage, in the case of both generation and consumption, is essential for the proper functioning of any aspect of our life.

One of the significant reasons for power loss all over the world is electricity theft\cite{powerquality}\cite{theft}. Electricity theft incurs substantial economic loss. It also accounts for various other issues like voltage imbalance, overload resulting in poor quality. The non-paying consumer tends to consume more power than expected. This often exceeds the nominal value and causes power quality problems in the grid. It also affects the generation unit and may cause damage to the appliances in place. Electrical faults on transmission lines and faults in electrical machines also contribute to power loss\cite{fault}. Faults can cause a huge load losses. A fault that causes a small voltage dip can also be hazardous to the grid and other appliances as it deviates the circuit from synchronisation.

Hence, the prevention of electricity thefts and faults becomes very important, which is only possible if their correct location is known. Finding the fault itself is very burdensome. It demands a considerable amount of manual labour and time as it may require examination of very long transmission lines (maybe tens of kilometres) or a large number of machines in a station. Therefore, a system that can detect possible points of theft and faults automatically can help to enhance power quality and economy of a region.

Moreover, all other physical quantities, that are measured using various sensors are indirectly obtained from voltage. Therefore, voltage data needs to be stored efficiently, so that it can be analysed for the efficient functioning of real-world systems. Hence, a system that can store voltage data efficiently is of utmost importance.

IoT can help to overcome the issues mentioned above. Small electronic devices can be used to measure the voltage values at various locations. The devices would be connected over a network and can transfer data over the network without any human intervention. A large amount of real-time data is obtained from IoT, which can be used for analysis. Therefore, the efficient storage of data collected from IoT is necessary.

A basic design choice for data storage would be Client-Server Model. Multiple clients would acquire voltage value using the voltage sensing devices and send it to the server, which would store and analyse the data. This model fails to work when the server shuts down for any reason. Also, there are lots of security concerns in this model.

A possible solution to overcome the issues mentioned above is the decentralisation of the system. A decentralised system is a system in which every node has an equal role to play. There is no classification of nodes into server and clients in this system. Blockchain and IPFS are two decentralised systems, which complement each other’s functionalities and together form a robust data storage system. Blockchain is a distributed ledger in which data is stored in all the nodes in the form of transactions. IPFS is a distributed file system. Therefore, in the decentralised model IPFS stores voltage data in files in a distributed manner and Blockchain keeps track of all the files being uploaded. Strong hashing and decentralisation make this a very good choice for voltage monitoring.

In this paper, we propose a centralised model and a decentralised model for voltage monitoring. The paper presents how these two models help to achieve the goal of Smart Voltage Monitoring. The rest of the paper has been outlined as follows:

Section II outlines the research conducted previously in the related fields. Section III addresses the challenges involved in voltage data measurement and storage. Section IV analyses the working and features of the centralised model thoroughly. Section V illustrates the decentralised model elaborately. Section VI compares the implementation of both the systems and the results obtained from the implementation. Section VII presents ideas for future development in the proposed models. Section VIII concludes the paper.
\section{Related Work}\label{sec:related}

In this section, we give an overview of the existing technologies that are related to voltage monitoring and IoT data storage. In \cite{svcms}, a method for measuring voltage is introduced. An Arduino is used to measure the data which is sent to an Android software using a Bluetooth module. This module, however, has many restrictions as Bluetooth has limited range and this may not be useful for continuous monitoring. In \cite{pmu1, pmu2, pmu3, pmu4}, various efficient techniques have been proposed to find the location of the fault using \emph{phasor measurement units(PMUs)}. But these papers don't address the storage of enormous voltage data produced. Moreover, PMUs are expensive. One possible cost-effective solution to monitor voltage and store voltage data efficiently is by integrating IoT. 

Several systems have been designed for IoT data storage and management for other spheres such as supply chain and healthcare. In \cite{jiang2014iot}, the IoT data obtained from RFID is stored in a cloud-computing platform. While in \cite{fu2018secure}, industrial IoT data has been stored by integrating fog computing and cloud computing. 
The authors in \cite{wang2018secure} have proposed solutions such as confidential data collection and confidential data storage in cloud-assisted IoT by taking privacy and security concerns into considerations. In recent years, blockchain has disrupted many industries because of its unique features such as decentralisation, security and immutability. This has led to its integration with IoT for efficient data management. But since the IoT data is abundant, they cannot be stored directly on blockchain as it would incur large fees in the form of a transaction fee. Therefore, the data is stored off-chain, and the hashes or the keys of the IoT data files or database are stored in the blockchain. In \cite{zhu2019controllable}, documents are stored in a cloud computing environment, and the corresponding hash of the file is stored in the blockchain. The changes made to any document can be tracked using blockchain. Also, there's a Trusted Authority (TA) in the network that administers all the processes in the system. 
While in the case of \cite{ayoade2018decentralized}, the data is stored in a Trusted Computing Environment - Intel SGX. The data is encrypted and stored into the trusted computing environment, and the corresponding hash of the file is stored in the blockchain. Whenever a third party needs some data, he has to request the owner of the data for decryption. He can then check the integrity of the file using the hash of the file. 
Most of the systems mentioned above for IoT data storage depend on a third-party \cite{jiang2014iot, fu2018secure, wang2018secure, zhu2019controllable} or make use of expensive hardware \cite{ayoade2018decentralized}. Thus, these systems are vulnerable and expensive. Aiming to overcome the issues mentioned above, a system was proposed in \cite{ali2017iot} which used Blockchain and IPFS. This increases the privacy of data significantly and makes the system independent from a third-party. These are some of the systems that have been built for IoT data storage.

But no specific system has been proposed or developed that implements both voltage monitoring and data storage efficiently. Therefore, a system that can store the voltage data efficiently and detect faults and theft issues is the need of the hour.

\section{Challenges Involved In Voltage Monitoring}\label{sec:challenges}
This section describes the major hurdles faced while attempting to monitor voltage.

\subsection{Voltage Measurement}\label{sub:sec:measurement}
Measuring the voltage manually at different sites would require coverage of large distances and involve the recording of a large amount of data in a short period.
Moreover, more human intervention makes the process time consuming and uneconomical. It is also dangerous for humans to measure the voltage at sites of high voltage. Hence, measuring voltage manually becomes impractical. This calls for greater automation in voltage measurement.

\subsection{Data Storage}\label{sub:sec:storage}
Storage of voltage data involves multiple complicacy. Storing the voltage data manually is close to impossible because the amount and rate of data generated are enormous. This calls for the automation of the process. Using a single machine to solve this issue would make the machine bulky and expensive. Hence, a cost-effective method needs to be developed such that the devices used are not a bulky and large amount of data can be stored efficiently.

\subsection{Security}\label{sub:sec:security}
If voltage data is not appropriately protected, anyone can access it illegally. From the voltage data, one can obtain sensitive information such as the voltage consumption of a house, which can convey the devices being used in a household. There also lies the issue of data being forged by some party. Therefore, the voltage data needs to be stored in a much more secure manner.

To overcome the challenges mentioned above in this section, we have proposed two solutions, namely VoltStar and VoltChain. Both of them help to monitor voltage efficiently with minimum human efforts. They only differ in the way voltage data is stored and analysed. The first solution, VoltStar, stores data in a centralised manner while the other, VoltChain, stores data in a decentralised way. Both of them are capable of detecting spots of electrical faults and thefts. They can also be used to keep track of various physical quantities like temperature and humidity by using respective sensors. Each of the solutions has been explained thoroughly in the paper. The complete working of each system and their features have been explained elaborately in the next two sections.

\section{VoltStar: A Centralised Voltage Monitoring System}\label{voltchain}
VoltStar is a voltage monitoring system that ensures continuous storage and analysis of voltage data in a centralised manner. 
In this section, the architecture, the functioning and various features of the centralized model have been described.

\begin{figure}[H]
\centerline{\includegraphics[scale=0.32]{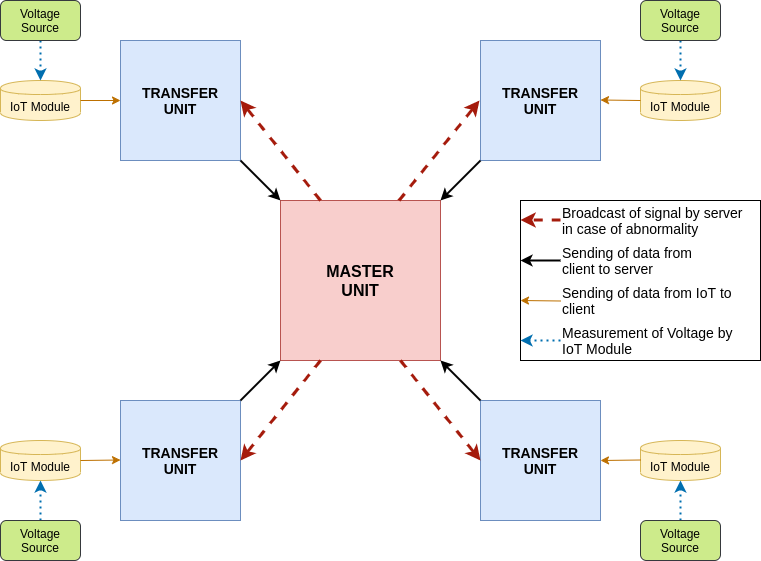}}
\caption{VoltStar: the centralised voltage monitoring system.}
\label{fig}
\end{figure}

\subsection{System Architecture}\label{sub:sec:voltstar-archi}
VoltStar is based on client-server\cite{oluwatosin2014client} network topology. Four components constitute this system - (1) Voltage Source; (2) IoT Module; (3) Transfer Unit; (4) Master Unit.

\textit{Voltage Source}: This is the point where the voltage has to be measured. It can be any point on grid or any machine in an industry.

\textit{IoT Module}: IoT Module is responsible for measuring the voltage at the voltage source. The IoT module can be built using basic electronic tools like Arduino, transformers and voltage dividers.

\textit{Transfer Unit (TU)}: Each IoT module has a personal computer(PC) connected to it. These PCs are the Transfer Units. They act as the interface between the IoT Module and the Master Unit.

\textit{Master Unit (MU)}: Master Unit is the control center of the network. It is responsible for storing and analysing all the data sent by each unit. Thus, MU reduces the load on the TUs significantly.

 Fig. 1 depicts the working of VoltStar. The TUs are all connected to the server within a common network, which can be wired or wireless. There is no interconnection between the TUs. The TUs acquire the measured voltage data from the IoT modules, pack the data into files, encrypt the data, and then send them to the MU. On receiving the files, the MU decrypts and analyses the files. If the MU detects some anomaly in any of the units, then it broadcasts a signal in the network to represent the location of the anomaly, so that whole of the region is notified and the issue is resolved as soon as possible. After analysis, the MU encrypts the files and keeps the files in its local storage. Encryption is meant to protect the voltage data. The MU needs to ensure that the decryption keys are kept secret.

The architecture of VoltStar is simple and hence, easy to implement. Since all the data is stored and processed in the MU, high computational power is needed in only one unit. However, VoltStar suffers from few limitations. If for some reason, the MU shuts down for some time, the whole network would collapse. Entire data of a network is contained only in the MU. Therefore, it is very easy for anyone to access the data illegally, once the decryption keys are obtained. This also increases the chances of breach of privacy. It is also simple to forge the data because the whole of the data is stored in the server, which may lead to wrong analysis.

\section{VoltChain: A Blockchain-based Decentralised Voltage Monitoring System}
VoltChain, a decentralised system, is designed aiming to overcome the limitations of VoltStar. A decentralised system does not have a central entity like a server. The data in the network is stored in a distributed manner. If a peer needs some data of another peer, it requests data directly from the peer that owns the data, not from a third-party. This section provides the necessary background which forms the base for our proposed decentralised solution. Then, it presents an elaborate description of the architecture and features of this system.

\subsection{Background}\label{AA}
Two technologies: (1) Blockchain; (2) IPFS form the building blocks of this system. Both of them have become increasingly popular in recent years because of the promising results that they have shown in various fields. Blockchain is a distributed ledger which helps to store data in a decentralised manner. IPFS is a distributed file system which helps to store files in a distributed manner and also facilitates file sharing.

\subsubsection{Blockchain}\label{AA}
Blockchain is a distributed database or a public ledger of transactions\cite{crosby2016blockchain}. Data in the blockchain is stored in the form of transactions. In a Blockchain network, every peer has a copy of the ledger. When any task is performed in the network, a transaction is created corresponding to it. Transactions contain information about the events occurring in the blockchain network. Once the transactions are added in the form of a \emph{block} to the blockchain network, every. For adding a block to the network, the block needs to be \emph{mined}. Mining in blockchain refers to completing a resource-consuming task to validate and add a block to the chain. This is the Proof-of-Work (PoW) consensus protocol, where the work refers to mining which involves generating a particular hash, which is challenging to produce. The 'work' deters denial-of-service-attacks and other service abuses such as spams on a network. There are other consensus protocols such as Proof-of-Stake (PoS), Proof-of-Capacity, but they are not as prevalent as PoW. The most common blockchain platforms: Bitcoin and Ethereum, are built on PoW. For making a transaction in the blockchain, one has to pay some price depending on the size of the data, like paying the Gas price in Ethereum or the Bitcoin Transaction Fee in case of Bitcoin. Therefore, storing real-time data directly on the blockchain is not possible as it would involve a large number of transactions, thereby incurring a large transaction fee. One way is to store the real-time data in files and then, storing the files in the blockchain. But, this would also incur large transaction fee as the size of files is much more than raw data. This is where IPFS comes into the picture. 

\subsubsection{IPFS}\label{AA}

IPFS is a distributed file storage system where files are stored in a decentralised manner\cite{benet2014ipfs}. Once a file is \emph{added} to the IPFS network, a hash is generated from the content of the file. The hash changes even if the slightest of changes are made to the content of the file, which makes IPFS immutable. Therefore, IPFS is a content-based system, i.e., the files can be traced in the system based on their content. It does not matter where the files are located (it is not address-based). Any peer can obtain a whole file of a second peer from IPFS if the peer owns the hash of the file. Hence, an efficient medium for hash-sharing is essential to work with IPFS. This is where blockchain helps. Blockchain stores the hashes of the file which \emph{added} into the network so that anyone in the network can \emph{get} it. There is no concept of any central entity like a server in IPFS too. IPFS is built on multiple proven software protocols. (1) Git, (2) BitTorrent, (3) Distributed Hash Tables(DHTs), (4) Self-Certified File Systems. These protocols together make IPFS the best among state of the art distributed file systems. Hence, Blockchain and IPFS together provide a decentralised, secure and more transparent data management system.

Blockchain and IPFS provide unique functionalities which have led to their obvious choice for the development of the decentralised model. \emph{Decentralisation} is the heart of blockchain and IPFS, which improves the performance of the system manifold and reduces the cost. \emph{Cryptography} forms the spine of both the systems. It is very difficult to obtain the actual file from its hash. Besides, a small tweak made to any data in a file would change the hash of the file completely. Hence, the files stored in IPFS are immutable. Furthermore, hashing of information makes the information in blockchain tamper-proof, thereby, making the data immutable. In blockchain, \emph{Smart Contracts} are computer programs that automatically execute the terms of a contract if certain conditions are met. Smart Contracts make the data management and process execution much more efficient and easier in the blockchain network. Next, we illustrate how these two technologies help us to build an effective system to accomplish our aim of monitoring voltage.





\subsection{System Architecture}\label{AA}
VoltChain is based on peer-to-peer architecture. VoltChain consists of three main components - (1) Voltage Source; (2) IoT Module; (3) Processing Unit. The Voltage Sources and the IoT Modules have exactly the same application as they had in the case of VoltStar.

\textit{Processing Unit (PU)}: In this model, each IoT module is connected to a Processing Unit, which is a PC. Each PU plays an equal role in this system in terms of storage and analysis. All the PUs are connected over two networks - a public blockchain and the IPFS. PUs acquire the voltage data from the IoT modules, put the data into files and upload these files into IPFS. The hash generated from IPFS is stored in the public blockchain in the form of transaction. Using the hash of the file, any PU can obtain any file uploaded into the network. At any point of time, the network has one \emph{Active} PU and rest are \emph{Dormant} PUs. All the PUs upload the voltage data files into IPFS and update the blockchain after uploading files into IPFS. Active PU performs additional tasks. It analyses the voltage data for a while to detect any anomaly in the network. To achieve this objective, the Active PU downloads all the files in IPFS using the hash uploaded registered in the public blockchain. If the Active PU detects any abnormal set of values in any unit, then it broadcasts a signal among all the units for quick resolution of the issue. Each PU has to perform the tasks of an Active PU for a certain amount of time in a day. This ensures even distribution of burden in the network. If the number of PUs is vast, then the system may need to have multiple Active PUs to analyse the network. This is how VoltChain functions to monitor the voltage of a region. The working of VoltChain is shown below in the diagram.

\begin{figure}[htbp]
\centerline{\includegraphics[scale=0.31]{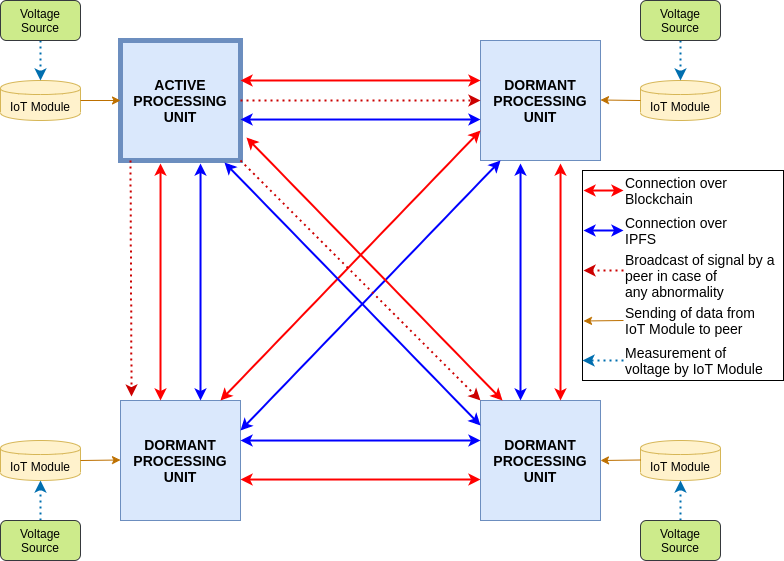}}
\caption{VoltChain: decentralised system network topology.}
\label{fig}
\end{figure}

Shutting down of one unit does not affect the whole system. Therefore, this system is fault tolerant. The foremost purpose served by this system is that it increases the data security manifold. Blockchain was integrated because the privacy and security of voltage data is enhanced\cite{blocksecurity1}. Blockchain also takes care of the access control of IoT data\cite{blocksecurity}. Data cannot be accessed illegally because of the secure hashing of the data. The integration of IPFS with blockchain improved the privacy even more\cite{ali2017iot}. Once the data is uploaded into the network, it cannot be tampered with. Hashing also makes it extremely difficult to obtain a file of another peer. There is transparency in the network. Anyone can keep monitor the changes occurring in the system.  

Just like VoltStar, VoltChain has its own limitations. Each of the peers needs to be computationally very powerful to make this architecture possible. This system architecture is complex and is based on relatively newer technologies. Hence, it is difficult to implement.

\section {Implementation And Results}\label{AA}
In this section, we present the implementation and evaluation of VoltStar and VoltChain.

\subsection{Implementation}\label{AA}
Here, we present how VoltStar and VoltChain were implemented. For both the models, the method used for data generation and file sharing have been described.
\subsubsection{VoltStar} VoltStar was implemented using Python 3.7 .

\emph{Data Generation: } In the TUs, to simulate the IoT data generation, virtual serial ports have been used. Virtual serial ports are generated in pairs, among which one is used for feeding data, and the other is used to read the fed data. Voltage data is fed at one of the ports. The baud rate of the serial was taken as 9600(this can be changed by the user). The voltage data was generated using the sine function of \emph{NumPy} module and by adding a random error to the value. Each cycle is sampled into 200 steps. The frequency of the sinusoidal function taken for the implementation is 50Hz (the frequency in grids all over India). The voltage values obtained and the timestamp of the time at which the data is obtained are stored in files of \emph{.csv} format. Data was generated in such fashion to get as close as possible to the way data would be generated by an IoT device connected to a voltage source.

\emph{File Sharing: }The number of TUs to be used is to be decided by the user. The MU listens on two ports - one is used for file sharing and the other is essential for broadcast during fault detection. The file generation and file sharing occurs simultaneously in a TU. In each TU, there are two tables - one table maintains the record of files being generated and the other keeps track of the files being sent. The MU analyses the files received. After analysis, the MU encrypts the files and stores it in its local storage. If a fault is detected, then the MU broadcasts in the other port. This ensures that the data storage remains unaffected even after a fault.

\subsubsection{VoltChain } This system was implemented using Ethereum\cite{wood2014ethereum}, ipfs v0.4.21 and Python 3.7(for interaction with blockchain and IPFS). Ethereum is a decentralized open source blockchain featuring smart contract functionality. Ether is the cryptocurrency used on Ethereum.

\emph{Data Generation: }The data generated in PUs was done exactly the same way as it was generated in the TUs in case of VoltStar.

\emph{File Sharing: } A private blockchain was used and ran \emph{ipfs daemon} is run to upload the files. The Smart Contract was written in Solidity, and Truffle \footnote{https://www.trufflesuite.com/truffle} was used for development. The web3Py module is used for establishing interaction between python and Smart Contract. For uploading files into IPFS, the \emph{ipfs daemon} has to be run throughout the process, and we used ipfs-api module for uploading files to IPFS from python. After the files are generated, each PU uploads the files into IPFS and the hash generated is stored into the blockchain. The \emph{Active} PU analyses the files. If it detects an anomaly, it sends an signal on the signal via a smart contract function.

\subsection{Evaluation}\label{sub:sec:evaluation}
The evaluation section has been subdivided into performance component and security analysis.
\subsubsection{Performance: }\label{sub:sub:sec:perfomance}Here, the setup of test environment has been described. The important implications derived from the tests have been discussed. The time required for file sharing and anomaly detection have been analysed because these are the metrics that would decide how quickly the fault would be detected. For VoltChain, the gas required for registering transactions has also been studied. The file distribution was done on a local network which would have lower latency compared to a live network. However, the trend of the graphs would remain the same even for a live network. The fault detection and gas consumption analysis would remain unaffected. All the metrics, i.e., file sharing, fault detection and gas consumption have been studied separately.

\emph{VoltStar: }This system was tested on an HP Pavilion - 15-cc134tx with 16GB of memory and an Intel i7 processor, on Ubuntu 18.04.4. For testing purposes, 10 TUs and a MU were considered. Each TU had to send ten files to the MU. The MU received files from all the TUs at the same time. The time needed to send files from the TUs to the MU over the socket was evaluated. The size of the files was varied by changing the number of rows in the voltage data files. The time required to detect the anomaly in the voltage data has also been evaluated. The RMS of voltage is calculated for 10 cycles and then compared with a preset threshold for the detection. We have considered the worst case for the evaluation, i.e., the anomaly occurs at the end of the file. 

\emph{VoltChain: }This system was also tested on the same testing environment. For setting up a blockchain environment on a single machine, we used Ganache \footnote{https://www.trufflesuite.com/ganache} which provides 10 accounts with 100 ethers each. The \emph{ipfs daemon} listened on \emph{localhost}. For testing purposes, 10 PUs were considered. Each PU had to upload 10 files to IPFS, and the PU also had to put the hash and name of the file into the blockchain. Then, the average time required to upload a file into IPFS and making a transaction into blockchain is calculated. The sum of these two yields the total time required to upload files into the network. The number of hashes being stored per transaction was varied to find the total gas requirement for registering hashes into blockchain. The results obtained in the implementation of VoltStar for fault detection also hold true for VoltChain.

\begin{figure}[H]
\centerline{\includegraphics[scale=0.40]{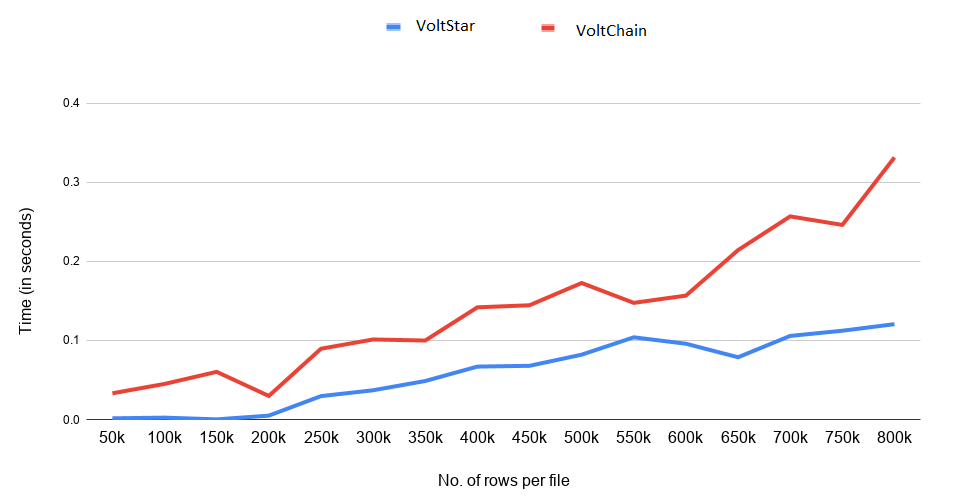}}
\caption{Comparison of time taken to transfer files for both systems }
\label{chart1}
\end{figure}

Here, we have presented the results obtained based on the analysis done on each system. 
Now, the results obtained from the from the testing have been analysed. Fig. \ref{chart1} represents the time taken for transfer of file with change in size. It is evident from the figure that the transfer of data is faster in VoltStar than VoltChain. On the other hand, VoltChain enhances security and privacy as described in Section \ref{voltchain}. Hence, this leads to an interesting trade-off between the security of data and the rate of data transfer. 

\begin{figure}[H]
\centerline{\includegraphics[scale=0.48]{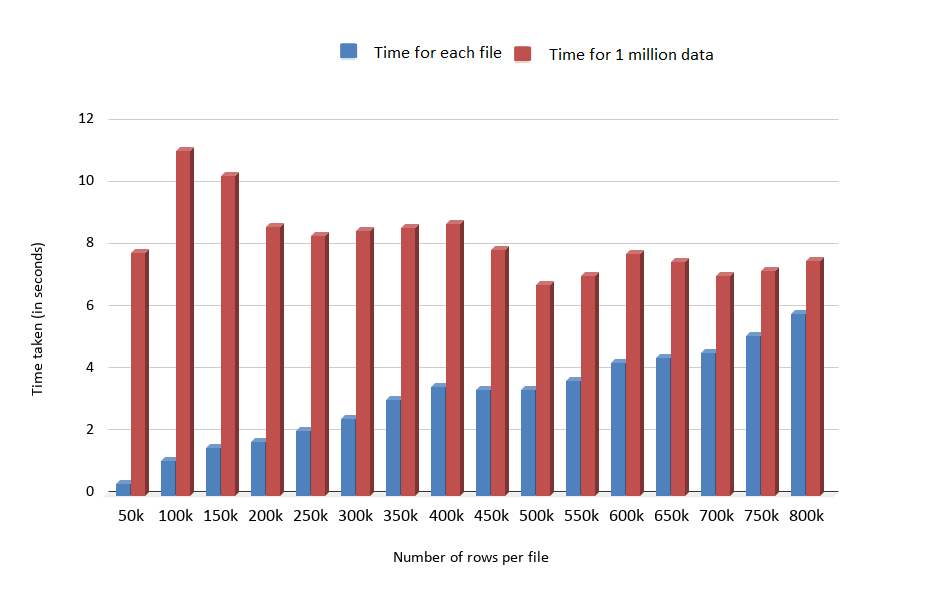}}
\caption{Time taken by the system to detect faults}
\label{chart2}
\end{figure}

Fig. \ref{chart2} represents the time taken to detect a fault. The blue columns represent the time required when there is a fault at the end of the file of that size. The red columns represent the time required to detect faults which occur at the end 1 million voltage data points with each file of having the number of data rows represented in the x-axis. The time required to find fault in a single file increases with the increase in voltage data. For detection of faults at the end of 1 million data points, the larger files perform better. The minimum time taken occurs when the file size is almost half of the total data that needs to be analysed. The time taken to detect faults in all the cases is small, which ensures early detection of faults in a system.

\begin{figure}[htbp]
\centerline{\includegraphics[scale=0.45]{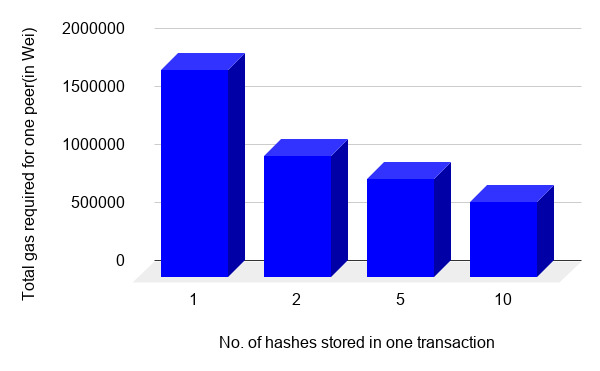}}
\caption{Amount of gas required with change in number of hashes being stored per transaction}
\label{chart3}
\end{figure}

When the number of hashes stored per transaction is less, the number of transactions is more. When more hashes are stored per transaction, more data is stored per transaction, which in turn incurs higher gas price per transaction. Hence, with the increase in the number of hashes stored per transaction, there's a conflicting effect of the increase in the size of data and a decrease in the number of transactions. It is clear from Fig. \ref{chart3} from the figure above that the effect of the decrease in the number of transaction is far more than the size of data stored per transaction. Therefore, storing multiple hashes per transaction is an excellent design choice. Retrieving hashes in any of the cases won't cost any ether because the required smart contract function would be of type \emph{view}.

\subsubsection{Security Analysis}

\begin{table}[htbp]
\begin{center}
\caption{Gas Price incurred to enhance security}
\label{tab:my_label}
\begin{tabular}{ |c|c| } 
\hline
Number of peers & Gas Price(in Wei) \\
\hline
0 & 133357 \\ 
1 & 153595 \\ 
2 & 174312 \\ 
3 & 194558 \\
4 & 214848 \\
5 & 235158 \\
6 & 214848 \\
7 & 194558 \\
8 & 174312 \\
9 & 153595 \\
10 & 133357\\

\hline
\end{tabular}
\end{center}
\end{table}

VoltChain makes the data sharing more secure. Hence, here we study the effect of security on performance. In TABLE \ref{tab:my_label}, we present the gas incurred in adding security to our decentralised. The left column represents the number of peers in the network who can acquire the IPFS hash of files. With increase in number of peers who can access a file, the gas required increases initially and then, decreases after reaching a peak. This is because when the number of peers that can access the hash is more than the half of the number of peers in the network, then it is better to store the IDs that cannot access the hash. When, the file is accessible to more than 50\% of the members of the network, then the level of security decreases. Hence, better the security of a file, more is the gas incurred. All the results furnish essential design patterns. The system can be designed based on the desired levels of security and speed using the results obtained from here.

\section{Future Work}\label{sec:future}
In this section, we have presented the future developments for both the systems that would make them more user-friendly and secure. 

\subsection{Dynamic Encryption}\label{sub:sec:dynamic}
Using different encryption schemes for different files would surely improve the security but would also increase the time taken to analyse. In both the models, the units can use dynamic encryption both while sending and storing files.

\subsection{Defence against network attacks}\label{sub:sec:dataflow}
VoltStar and VoltChain are still not entirely resistant to various network attacks. Methods described in \cite{dos, sybil, sniffing, dnsspoof} can be systems to build defence systems against Denial of Service, Sybil, Sniffing and DNS Spoofing attacks respectively. This implementation is necessary for making the data secure.

\subsection{Mobile-App Integration}\label{AA}
A mobile app that can represent the voltage behaviour of any region to anyone in the network would be beneficial for this system.
Such an application would enhance the detection of abnormalities as it would make the technology accessible for everyone but it would also introduce privacy issues.

\subsection{Use of light clients in Decentralised System}\label{AA}
In case of VoltChain, each unit needs to be computationally powerful because of the high computational power required for the blockchain and IPFS. This issue can be overcome if we use their \emph{light client}\cite{bunz2019flyclient} version. Light client version would help to use blockchain and IPFS without having to run their full versions on a local machine. However, the light client versions are not ready full-scale integration but are in rapid development.

\section{Conclusion}

This paper proposes two effective methods that can be used for voltage monitoring. 
The centralised solution is fast but is more vulnerable to attacks. The decentralised solution enhances security but is relatively slow. Therefore, it is up to the user to decide which system should be used.
Moreover, the characteristics of the network can be varied according to needs based on the results obtained in Section \ref{sub:sec:evaluation}. 
With the amount of data being generated increasing at a tremendous rate, the significance of privacy and security of data surging and more IoT devices becoming pervasive, the decentralised model becomes an attractive solution. If the ideas put up for future development are implemented, then it would lead to the establishment of a voltage monitoring system which would be efficient and resistant to attacks.

\bibliographystyle{ieeetr}
\bibliography{main}

\end{document}